\documentclass[preprintnumbers,amsmath,amssymb]{revtex4}
\usepackage{graphicx}
\usepackage{amsmath}
\usepackage{mathtools}

\newcommand{\comm}[1]{}
\newcommand{\bk}[2]{\langle\: #1\mid #2\:\rangle }

\begin{document}

\title{Space time symmetry in quantum mechanics}
\author{Zinkoo Yun}
 \email{semiro@uvic.ca}
\affiliation{Department of Physics and Astronomy University of Victoria, Canada}

%\date{\ October 29, 2013}

\begin{abstract}
New prescription to treat position and time equally in quantum mechanics is presented. Using this prescription, we could successfully derive some interesting formulae such as time-of-arrival for a free particle and quantum tunneling formula. The physical interpretation will be discussed.
\end{abstract}


\maketitle

\section{Introduction}
Though we can treat time and space symmetric way in relativity, in quantum mechanics the time seems different to other observables: It seems we don't have proper operator for time. A particle detected at one position can be detected at the same position at later time, namely, we encounter the difficulty on orthogonality and normalization and these two measurements do not commute each other. This non commuting property leads us to think about time-of-arrival which means  the time that a particle first arrives to a specific position. 

Allcock \cite{allcock} tried to build time-of-arrival eigenstates which are orthogonal each other for different time but could not define a consistent time-of-arrival. His study says that because we cannot absorb the particle in an arbitrarily short time, we cannot measure the time-of-arrival at any accuracy. Oppenheim \emph{et al}\cite{unruh} insist that using two level detector absorbing a particle in arbitrarily short period time, we can overcome this restrictions. However they found that the limitation on measuring the time-of-arrival with arbitrarily accuracy comes from the clock coupled to the trigger.
They show that if we couple the system to a clock to measure the time-of-arrival at which the particle arrives at specific position, then the accuracy of measurement is limited by
\[
\delta t_A > \hbar/E_k
\]
where $\delta t_A$ is the minimum uncertainty of measuring time of arrival and $E_k$  is the energy of clock.

One of interesting approaches to find the time-of-arrival operator has been studied by Grot  \emph{et al}.\cite{grot} They tried to develop the time-of-arrival operator for free non-relativistic particle by proper ordering of space time operator in Heisenberg picture, analogous to classical picture. But the eigenstates of time-of-arrival they calculated satisfying eigenvalue equation, did not satisfy the orthogonal condition for different times. They bypassed this difficulty by modifying the time-of-arrival operator so that in the classical limit would not reproduced the time-of-arrival exactly, but would reproduce a quantity arbitrary close to the time-of-arrival. 

In this article, I will not attempt to develop the time-of-arrival operator nor discuss about dynamical limitation on treating position and time of quantum mechanics in equal manner. Rather it will be focused on how we can put an equal footing on position and time in quantum mechanical evolution. Contrast to other approaches, I assumed that we cannot put an equal footing on both position and time simultaneously. That is, when we treat the position as a quantum operator, we have to treat the time as an evolution parameter. And when we treat the time as a quantum operator, we have to treat the position as an evolution parameter.
We will discuss how we can apply this prescription on quantum tunneling process.

\section{Prescription} 
In this section notational conventions will be defined in symmetrical way for both position and time. When the time is used as an evolution parameter (TEP), the position is used as a usual quantum observable. When the position is used as an  evolution parameter (PEP), the time is used as a usual quantum observable. We can specify any quantum states with one state vector and one evolution parameter.
\subsection{TEP}  
We denote the quantum state $\psi$ at time $t_1$ by
\begin{equation}\label{ligah1a} 
\mid \psi, \underline{t_1}\rangle
\end{equation} 
where the underline on $\underline{t_1}$ means that time is the evolution parameter. That is, the first one represents the quantum state and the second one stands for the evolution parameter. 
 We can express (\ref{ligah1a}) it shorter by 
\begin{equation}\label{ligah1b} 
\mid \psi_1\rangle \equiv \mid \psi, \underline{t_1}\rangle
\end{equation} 
 where the subscript 1 on $\psi$ means the state is of time $t_1$. 
The probability amplitude to find the position $x$ is then
\begin{equation}\label{ligah1c} 
\langle x \mid \psi, \underline{t_1}\rangle = \psi (x,\underline{t_1})
\end{equation} 
where the position $x$ is the usual quantum observable and the time $\underline{t_1}$ is the evolution parameter. Thus while the operation $\langle x\mid x_1\rangle$ is possible, the operation $\langle \underline{t}\mid \underline{t_1}\rangle$ is not possible because both $t$ and $t_1$ are just evolution parameters. By the same reason $\langle\underline{t}\mid \psi, \underline{t_1}\rangle=\psi(\underline{t},\underline{t_1})$ does not make sense. Since this represents the pure evolution process up to $t$, it must be denoted by
\begin{equation}\label{ligah1d} 
\langle\underline{t}\mid \psi, \underline{t_1}\rangle=\mid \psi, \underline{t}\rangle
\end{equation} 
\subsection{PEP}  
We denote the quantum state $\varphi$ at position $x_1$ by
\begin{equation}\label{ligah2a} 
\mid \varphi, \underline{x_1}\rangle
\end{equation} 
where the underline on $\underline{x_1}$ means that position is the evolution parameter. That is, the first one represents the quantum state and the second one stands for the evolution parameter. 
 We can express (\ref{ligah2a}) it shorter by 
\begin{equation}\label{ligah2b} 
\mid \varphi_1\rangle \equiv \mid \varphi, \underline{x_1}\rangle
\end{equation} 
 where the subscript 1 on $\varphi$ means the state is of position $x_1$. 
The probability amplitude to find the time $t$ is then
\begin{equation}\label{ligah2c} 
\langle t \mid \varphi, \underline{x_1}\rangle = \varphi (t,\underline{x_1})
\end{equation} 
where the time $t$ is the usual quantum observable and the position $\underline{x_1}$ is the evolution parameter. Thus while the operation $\langle t\mid t_1\rangle$ is possible, the operation $\langle \underline{x}\mid \underline{x_1}\rangle$ is not possible because both $x$ and $x_1$ are just evolution parameters. By the same reason $\langle\underline{x}\mid \varphi, \underline{x_1}\rangle=\varphi(\underline{x},\underline{x_1})$ does not make sense. Since this represents the pure evolution process up to $x$, it must be denoted by
\begin{equation}\label{ligah2d} 
\langle\underline{x}\mid \varphi, \underline{x_1}\rangle=\mid \varphi, \underline{x}\rangle
\end{equation}
\\ 

As we have seen, any quantum state is expressed by one state vector and one evolution parameter as $\mid \psi,\underline{t}\rangle$ or $\mid \varphi, \underline{x}\rangle$. We cannot specify a quantum state by $\mid \psi,x,\underline{t}\rangle$ (two state vectors) or by $\mid \psi,\underline{x},\underline{t}\rangle$ (two evolution parameters).

The rule is simple: The state vector $\psi$ does not operate to the evolution
parameter $\underline{e}$. So $\langle\underline{e}\mid\psi\rangle$ does not work.
Only two exceptions are $\langle \underline{t}\mid
E_n\rangle=e^{-iE_nt}$ and $\langle\underline{x} \mid
p_n\rangle=e^{ip_nx}$ where $E_n$, $p_n$ are the components of identity
operator $I=\sum_n\mid p_n\rangle\langle p_n\mid=\sum_n\mid E_n\rangle\langle
E_n\mid$. If $E$ is not a component of identity operator and  $\langle
\underline{t}\mid E \rangle=e^{-iEt}$ contributes only overall phase, then
it is physically meaningless. 

And two state vectors with different evolution
parameters do not \emph{directly} operate each other. For example, two state vectors
$\varphi$ and $\psi$ in
$\langle\varphi,\underline{t_2}\mid \psi,\underline{t_1}\rangle$ do not
\emph{directly} operate each other.
In this case, if needed, we may sandwich
\begin{equation}\label{ligah2d1} 
I=\sum_n\mid E_n\rangle\langle E_n\mid
\end{equation}
between $\underline{t_2}$ and  $\underline{t_1}$ or we may sandwich
\begin{equation}\label{ligah2d2} 
I=\sum_n\mid p_n\rangle\langle p_n\mid
\end{equation}
between $\underline{x_2}$ and  $\underline{x_1}$.
\footnote{ 
 Note in order to use new prescription, it is not good idea to sandwich (\ref{ligah2d1}) between $\b{x\tiny{2}}$ and  $\b{x\tiny{1}}$ / to sandwich (\ref{ligah2d2}) between $\b{t\tiny{2}}$ and  $\b{t\tiny{1}}$, because it can destroy the information about the Lagrangian of the system. 
}

\section{Expression in $E$ and $p$ basis} 
\subsection{TEP}  
We can express the quantum state $\psi$  at time $t_1$ by
\begin{equation}\label{ligah3a} 
\mid \psi_1,\underline{t_1}\rangle = \mid  E_n\rangle \langle E_n\mid \psi_1,\underline{t_1}\rangle
\end{equation} 
where the summation is assumed for repeated index $n$.
Thus the probability amplitude to find the state $\psi_2$ at $t_2$ is
\begin{equation}\label{ligah3b} 
\langle \psi_2,\underline{t_2}\mid \psi_1,\underline{t_1}\rangle =\langle \psi_2,\underline{t_2}\mid E_n\rangle\langle E_n\mid \psi_1,\underline{t_1}\rangle 
\end{equation} 
For example, $\psi_2=x_2$,
\begin{eqnarray*}
\langle x_2,\underline{t_2}\mid\psi_1,\underline{t_1}\rangle
&=&\langle x_2,\underline{t_2}\mid E_n\rangle\langle E_n\mid
\psi_1,\underline{t_1}\rangle\\
\langle x_2\mid\psi_2,\underline{t_2}\rangle &=& \langle x_2\mid
E_n\rangle\langle \underline{t_2}\mid E_n\rangle\langle
E_n\mid\psi_1\rangle\langle E_n\mid \underline{t_1}\rangle\\
\to \psi_2(x_2,\underline{t_2}) &=& \int \langle x_2\mid E\rangle
\psi_1(E)
e^{-iE(t_2-t_1)}{dE}
\end{eqnarray*}
where the subscripts 1 and 2 in $\psi$ mean the states $\psi$ is of $t_1$ and $t_2$, we set $\hbar=1$. 

\subsection{PEP}  
We can express the quantum state $\varphi$  at position $x_1$ by
\begin{equation}\label{ligah4a} 
\mid \varphi_1,\underline{x_1}\rangle = \mid p_n\rangle \langle p_n\mid \varphi_1,\underline{x_1}\rangle
\end{equation} 
The probability amplitude to find the state $\varphi_2$ at $x_2$ is
\begin{equation}\label{ligah4b} 
\langle \varphi_2,\underline{x_2}\mid \varphi_1,\underline{x_1}\rangle =\langle \varphi_2,\underline{x_2}\mid p_n\rangle\langle p_n\mid \varphi_1,\underline{x_1}\rangle 
\end{equation} 
For example, $\varphi_2=t_2$,
\begin{eqnarray}
\langle t_2,\underline{x_2}\mid\varphi_1,\underline{x_1}\rangle
&=&\langle t_2,\underline{x_2}\mid p_n\rangle\langle p_n\mid
\varphi_1,\underline{x_1}\rangle\nonumber\\
\langle t_2\mid\varphi_2,\underline{x_2}\rangle &=& \langle t_2\mid
p_n\rangle\langle \underline{x_2}\mid p_n\rangle\langle
p_n\mid\varphi_1\rangle\langle p_n\mid \underline{x_1}\rangle\nonumber\\
\to \varphi_2(t_2,\underline{x_2}) &=& \int \langle t_2\mid p\rangle \varphi_1(p)
e^{ip(x_2-x_1)}dp\label{ligah4b1}
\end{eqnarray}

\section{Expression in space time basis} 
\begin{figure}
\begin{center}
    \includegraphics[height=6cm]{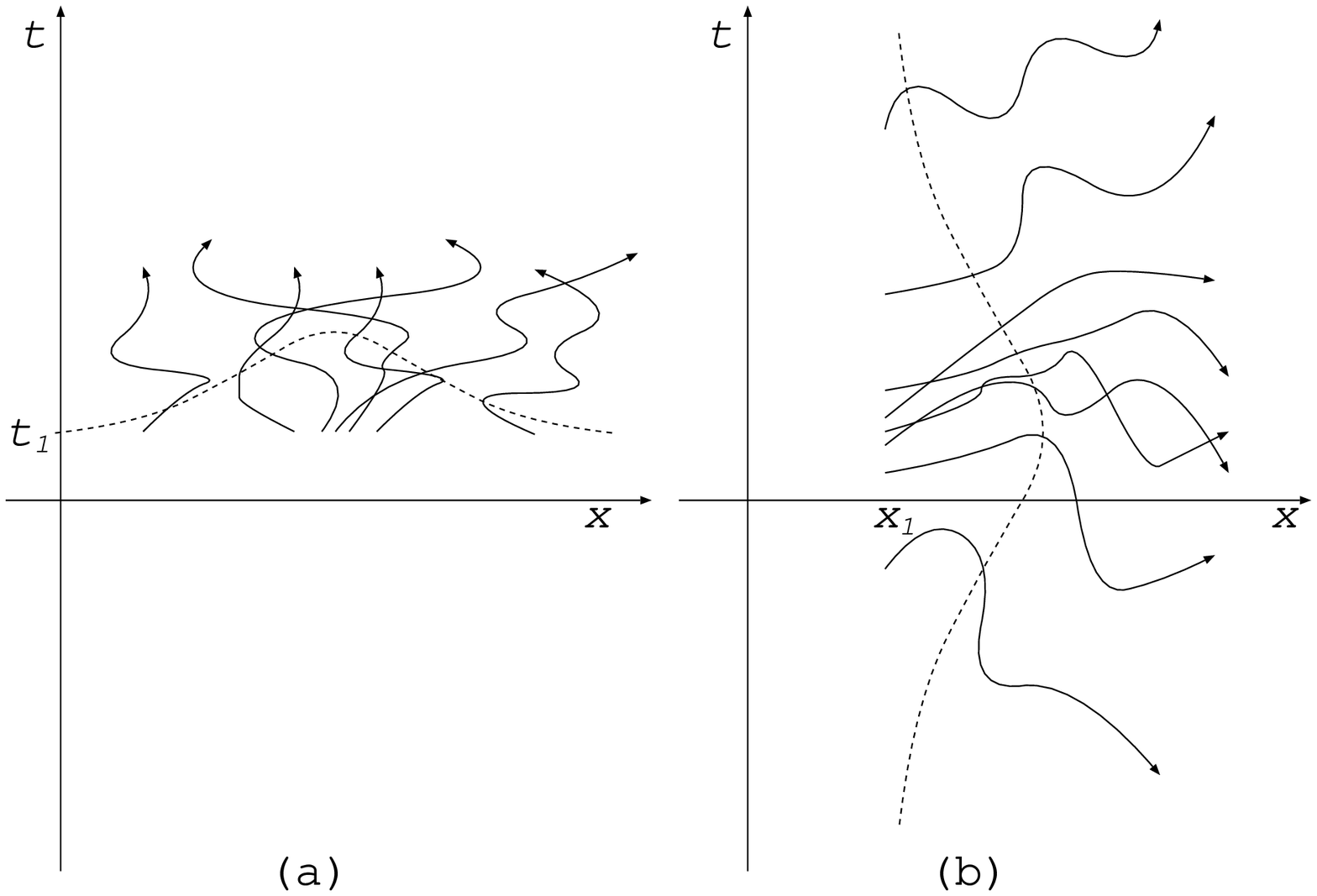}
  \end{center}
  \caption{(a) Illustration of (\ref{ligah5b}); (b) Illustration of (\ref{ligah6b})}
\label{tepeptun_fig1}
\end{figure}

\subsection{TEP}  
 In (\ref{ligah3a}) and (\ref{ligah3b}), let's use $\mid x_i,\underline{t_1}\rangle\langle x_i,\underline{t_1}\mid$ instead of $ \mid E_n\rangle\langle E_n\mid$.
\begin{eqnarray}
\mid\psi_1,\underline{t_1}\rangle &=& \mid x_i,\underline{t_1}\rangle\langle x_i,\underline{t_1}\mid \psi_1,\underline{t_1}\rangle=\int \mid x,\underline{t_1}\rangle\langle x\mid \psi_1\rangle dx \label{ligah5a} \\
&=&  \int \mid x,\underline{t_1}\rangle\langle x\mid p_n\rangle\langle p_n\mid \psi_1\rangle dx=\int\mid x,\underline{t_1}\rangle\frac{e^{ip_nx}}{\sqrt{2\pi}}\langle p_n\mid \psi_1\rangle dx\label{ligah5b}
\end{eqnarray}
For the momentum eigenstate $\psi_1=p_1$,
\begin{equation}\label{ligah5c} 
\mid p_1,\underline{t_1}\rangle=\int \mid x,\underline{t_1}\rangle\frac{e^{ip_1x}}{\sqrt{2\pi}}dx
\end{equation} 
 From (\ref{ligah5c}), we can see that a particle in momentum eigenstate evolving from time $\underline{t_1}$ starts its motion at all equally different positions. (\ref{ligah5b}) is illustrated in figure \ref{tepeptun_fig1}(a). We can check that if we are more certain about the momentum of a particle, we are less certain about the position of departure at time $\underline{t_1}$.   This is the fundamental meaning of position-momentum uncertainty relation.

\subsection{PEP}  
 In (\ref{ligah4a}) and (\ref{ligah4b}), let's use $\mid t_i,\underline{x_1}\rangle\langle t_i,\underline{x_1}\mid$ instead of $\mid p_n\rangle\langle p_n\mid$.
\begin{eqnarray}
\mid\varphi_1,\underline{x_1}\rangle &=&  \mid t_i,\underline{x_1}\rangle\langle t_i,\underline{x_1}\mid \varphi_1,\underline{x_1}\rangle=\int \mid t,\underline{x_1}\rangle\langle t\mid \varphi_1\rangle dt \label{ligah6a} \\
&=&\int \mid t,\underline{x_1}\rangle\langle t\mid E_n\rangle\langle E_n\mid \varphi_1\rangle dt=\int\mid t,\underline{x_1}\rangle\frac{e^{-iE_nt}}{\sqrt{2\pi}}\langle E_n\mid \varphi_1\rangle dt\label{ligah6b}
\end{eqnarray}
For the energy eigenstate $\varphi_1=E_1$,
\begin{equation}\label{ligah6c} 
\mid E_1,\underline{x_1}\rangle=\int \mid t,\underline{x_1}\rangle\frac{-e^{iE_1t}}{\sqrt{2\pi}}dt
\end{equation} 
From (\ref{ligah6c}), we can see that the particle in energy eigenstate evolving from the position $\underline{x_1}$ starts its motion at all equally different times. (\ref{ligah6b}) is illustrated in figure \ref{tepeptun_fig1}(b). We can check that if we are more certain about the energy of a particle, we are less certain about the time of departure at position $\underline{x_1}$.   This is the fundamental meaning of time-energy uncertainty relation.

\section{Quantum evolution in space time basis} 
\begin{figure}
\begin{center}
    \includegraphics[height=6cm]{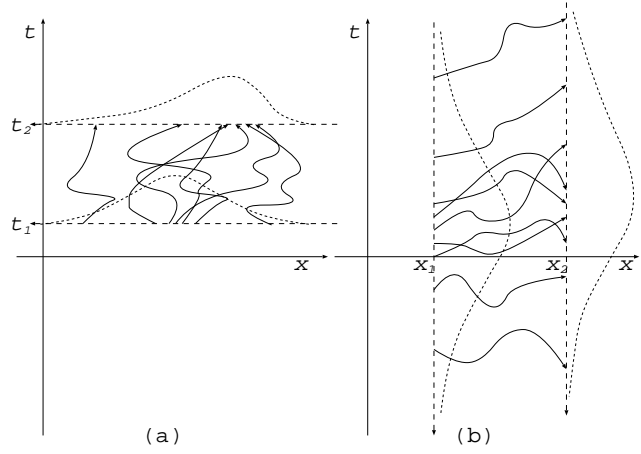}
  \end{center}
  \caption{(a) Illustration of (\ref{ligah7c}); (b) Illustration of (\ref{ligah8c})}
\label{tepeptun_fig2}
\end{figure}
\subsection{TEP}  
Let's find the expression for the probability amplitude of quantum state $\psi_1$ at $\underline{t_1}$ to be measured $\psi_2$ at $\underline{t_2}$.
\begin{eqnarray}
\langle \psi_2, \underline{t_2}\mid \psi_1,\underline{t_1}\rangle 
\comm{&=& \int \langle \psi_2,\underline{t_2}\mid x_2,\underline{t_2}\rangle\langle x_2,\underline{t_2}\mid \psi_1, \underline{t_1} \rangle dx_2 \nonumber\\}
&=& \int \underbrace{\langle \psi_2,\underline{t_2}\mid x_2,\underline{t_2}\rangle}_{\langle \psi_2\mid x_2\rangle}\langle x_2,\underline{t_2}\mid x_1,\underline{t_1}\rangle
 \underbrace{\langle x_1,\underline{t_1}\mid \psi_1, \underline{t_1}\rangle}_{\langle x_1\mid \psi_1\rangle}dx_1 dx_2\nonumber\\
&=&\iint \langle \psi_2\mid x_2\rangle\langle x_2,\underline{t_2}\mid x_1,\underline{t_1}\rangle
\langle x_1\mid\psi_1\rangle dx_1 dx_2 \label{ligah7c}
\end{eqnarray}
Equation (\ref{ligah7c}) is illustrated in figure \ref{tepeptun_fig2}(a).

For example
\begin{eqnarray}
\langle x_2,\underline{t_2}\mid \psi, \underline{t_1}\rangle &=&\iint \langle x_2\mid x_2'\rangle\langle x_2', \underline{t_2}\mid x_1,\underline{t_1}\rangle
\langle x_1\mid \psi_1\rangle dx_1 dx_2'\\
\langle x_2\mid \psi, \underline{t_2}\rangle &=& \int \langle x_2,\underline{t_2}\mid x_1,\underline{t_1}\rangle\langle x_1\mid \psi_1\rangle dx_1\\
\to \psi(x_2,\underline{t_2}) &=& \int \langle x_2,\underline{t_2}\mid x_1,\underline{t_1}\rangle \psi (x_1,\underline{t_1})dx_1\label{ligah7f}
\end{eqnarray}
\comm{
(\ref{ligah7f}) leads to Schrodinger's equation.
}

Take an another example
\begin{align}
\langle  p_2,\underline{t_2} \mid p_1,\underline{t_1}\rangle &= \iint \langle p_2\mid x_2\rangle\langle x_2,\underline{t_2}\mid x_1,\underline{t_1}\rangle\langle x_1\mid p_1\rangle dx_1 dx_2\nonumber\\
&= \iint \frac{e^{-ip_2x_2}}{\sqrt{2\pi}}\langle x_2,\underline{t_2}\mid x_1,\underline{t_1}\rangle \frac{e^{ip_1x_1}}{\sqrt{2\pi}} dx_1 dx_2 
\end{align}

\subsection{PEP}  
Let's find the expression for the probability amplitude of quantum state $\varphi_1$ at $\underline{x_1}$ to be measured $\varphi_2$ at $\underline{x_2}$.
\begin{eqnarray}
\langle \varphi_2, \underline{x_2}\mid \varphi_1,\underline{x_1}\rangle 
\comm{&=& \int \langle \varphi_2,\underline{x_2}\mid t_2,\underline{x_2}\rangle\langle t_2,\underline{x_2}\mid \varphi_1, \underline{x_1} \rangle dt_2 \nonumber\\}
&=& \int \underbrace{\langle \varphi_2,\underline{x_2}\mid t_2,\underline{x_2}\rangle}_{\langle \varphi_2\mid t_2\rangle}\langle t_2,\underline{x_2}\mid t_1,\underline{x_1}\rangle
\underbrace{\langle t_1,\underline{x_1}\mid \varphi_1, \underline{x_1}\rangle}_{\langle t_1\mid \varphi_1\rangle}dt_1 dt_2\nonumber\\
&=&\iint \langle \varphi_2\mid t_2\rangle\langle t_2,\underline{x_2}\mid t_1,\underline{x_1}\rangle
\langle t_1\mid\varphi_1\rangle dt_1 dt_2 \label{ligah8c}
\end{eqnarray}
Equation (\ref{ligah8c}) is illustrated in figure \ref{tepeptun_fig2}(b).

For example
\begin{eqnarray}
\langle t_2,\underline{x_2}\mid \varphi, \underline{x_1}\rangle &=&\iint \langle t_2\mid t_2'\rangle\langle t_2', \underline{x_2}\mid t_1,\underline{x_1}\rangle
\langle t_1\mid \varphi_1\rangle dt_1 dt_2'\\
\langle t_2\mid \varphi, \underline{x_2}\rangle &=& \int \langle t_2,\underline{x_2}\mid t_1,\underline{x_1}\rangle\langle t_1\mid \varphi_1\rangle dt_1\\
\to \varphi(t_2,\underline{x_2}) &=& \int \langle t_2,\underline{x_2}\mid t_1,\underline{x_1}\rangle \varphi (t_1,\underline{x_1})dt_1\label{ligah8f}
\end{eqnarray}
\comm{
Could (\ref{ligah8f}) lead to counter part of Schrodinger's equation?
}

Take an another example
\begin{align}
\langle  E_2,\underline{x_2}\mid E_1,\underline{x_1}\rangle &= \iint \langle E_2\mid t_2\rangle\langle t_2,\underline{x_2}\mid t_1,\underline{x_1}\rangle\langle t_1\mid E_1\rangle dt_1 dt_2\nonumber\\
&= \iint \frac{e^{iE_2t_2}}{\sqrt{2\pi}}\langle t_2,\underline{x_2}\mid t_1,\underline{x_1}\rangle \frac{e^{-iE_1t_1}}{\sqrt{2\pi}} dt_1 dt_2 \label{ligah8h}
\end{align}

\section{Orthogonality of position and time}
In this section it will be shown how to achieve $\bk{x}{x'}=\delta(x-x')$ in TEP and  $\bk{t}{t'}=\delta(t-t')$ in PEP.
In doing so, we will find the following expression for $\bk{E}{p}$.
\comm{The information about the system such as Lagrangian of the system is in $\bk{E}{p}$. In other words, we need to know about the system to express $\bk{E}{p}$.}
\begin{equation}\label{ligai1a}
\bk{E}{p}=\sqrt{\frac{E}{p_E}}\delta(\pm p_E-p) \qquad \textrm{in TEP}
\end{equation}
The range of energy is either $[m,\infty]$ or $[-m,-\infty]$. The range of momentum is $[-\infty,\infty]$.

\begin{equation}\label{ligai2a}
\bk{E}{p}=\sqrt{\frac{p}{E_p}}\delta(\pm E_p-E) \qquad \textrm{in PEP}
\end{equation}
The range of momentum is either $[0,\infty]$ or $[0,-\infty]$. The range of energy is $[-\infty,\infty]$ except $[-m,m]$.
Where $p_E\equiv \sqrt{E^2-m^2}$ and $E_p\equiv\sqrt{p^2+m^2}$ for a free particle.
Let's check this out.

\subsection{TEP}
In order to find out the expression of $\bk{E}{p}$, first verify the orthogonality of momentum, 
$\bk{p'}{p}=\int\bk{p'}{E}\bk{E}{p}dE=\delta(p'-p)$.
Since $EdE=p_Edp_E$, try $\bk{E}{p}=\sqrt{E/p_E}\delta(\pm p_E-p)$, where $\pm$ indicates that when we integrate over $p_E$,
we have to do it for both $+p_E$ and $-p_E$.
\begin{equation}\label{ligai1c}
\bk{p'}{p}=\int \sqrt{\frac{E}{p_E}}\delta(\pm p_E-p')\sqrt{\frac{E}{p_E}}\delta(\pm p_E-p)dE
\end{equation}
This is an odd function of $E$, thus if we integrate over $[-\infty,\infty]$, it turns out to be zero. We can fix it
by restricting $E$ to either $[m,\infty]$ or $[-m,-\infty]$. Then,
\begin{equation}\label{ligai1d}
\bk{p'}{p}=\int^\infty_m \frac{E}{p_E}\delta(\pm p_E-p')\delta(\pm p_E-p)dE
\end{equation}
The sign of $E$ does not specify the sign of $p_E$. Thus we have to count both positive and negative momentum cases.
\begin{align}
\bk{p'}{p} &=\int^\infty_0\delta(p_E-p')\delta(p_E-p)dp_E+\int^\infty_0\delta(p_E+p')\delta(p_E+p)dp_E\\
&=\int^\infty_0\delta(p_E-p')\delta(p_E-p)dp_E+\int^{-\infty}_0\delta(-u+p')\delta(-u+p)d(-u)\\
&=\int^\infty_{-\infty}\delta(p_E-p')\delta(p_E-p)dp_E=\delta(p'-p)
\end{align}
\\

(\ref{ligai1a}) with $E$  either $[m,\infty]$ or $[-m,-\infty]$ ensures the orthogonality of position, 
$\bk{x}{x'}=\delta(x-x')$:
\begin{equation}\label{ligai1e}
\bk{x}{E}=\int\bk{x}{p}\bk{p}{E}dp=\frac{1}{\sqrt{2\pi}}\int e^{ipx}\sqrt{\frac{E}{p_E}}\delta(\pm p_E-p)dp
=\frac{1}{\sqrt{2\pi}}\sqrt{\frac{E}{p_E}}e^{\pm i p_E x}
\end{equation}
\begin{align}
\bk{x}{x'} &=\int^\infty_m\bk{x}{E}\bk{E}{x'}dE=\frac{1}{2\pi}\int^\infty_m\frac{E}{p_E}e^{\pm ip_E(x-x')}dE\\
\comm{&=\frac{1}{2\pi}\int^\infty_0\frac{E}{p_E}e^{ip_E(x-x')}\frac{p_E}{E}dp_E
+\frac{1}{2\pi}\int^\infty_0\frac{E}{p_E}e^{-ip_E(x-x')}\frac{p_E}{E}dp_E\\
&=\frac{1}{2\pi}\int^\infty_0 e^{ip_E(x-x')}dp_E+\frac{1}{2\pi}\int^{-\infty}_0 e^{iu(x-x')}d(-u)\\}
&=\frac{1}{2\pi}\int^\infty_{-\infty} e^{ip_E(x-x')}dp_E=\delta(x-x')
\end{align}
If we did not restrict $E$ to either positive or negative values, we couldn't have $\bk{x}{x'}=\delta(x-x')$.
This is expected because the negative energy particle comes backward in time to be detected at another position
at the same time it has already been detected.

\subsection{PEP}
In order to find out the expression of $\bk{E}{p}$, first verify the orthogonality of energy, 
$\bk{E'}{E}=\int\bk{E'}{p}\bk{p}{E}dp=\delta(E'-E)$.
Since $EdE=p_Edp_E$, try $\bk{E}{p}=\sqrt{p/E_p}\delta(\pm E_p-E)$. 
\begin{equation}\label{ligai2c}
\bk{E'}{E}=\int \sqrt{\frac{p}{E_p}}\delta(\pm E_p-E')\sqrt{\frac{p}{E_p}}\delta(\pm E_p-E)dp
\end{equation}
This is an odd function of $p$, thus if we integrate over $[-\infty,\infty]$, it turns out to be zero. We can fix it
by restricting $p$ to either $[0,\infty]$ or $[0,-\infty]$.\comm{but not $[-\infty,0]$}
\footnote{We may fix this problem by making the integrand to  an even function of $p$ (i.e., $p\to|p|$). But
as explained at the appendix \ref{epeven}, this cause another problem.} Then,
\begin{equation}\label{ligai2d}
\bk{E'}{E}=\int^\infty_0 \frac{p}{E_p}\delta(\pm E_p-E')\delta(\pm E_p-E)dp
\end{equation}
The sign of $p$ does not specify the sign of $E_p$. Thus we have to count both positive and negative energy cases.
\begin{align}
\bk{E'}{E} &=\int^\infty_0\delta(E_p-E')\delta(E_p-E)dE_p+\int^\infty_0\delta(E_p+E')\delta(E_p+E)dE_p\\
&=\int^\infty_0\delta(E_p-E')\delta(E_p-E)dE_p+\int^{-\infty}_0\delta(-u+E')\delta(-u+E)d(-u)\\
&=\int^\infty_{-\infty}\delta(E_p-E')\delta(E_p-E)dE_p=\delta(E'-E)
\end{align}
\\

(\ref{ligai2a}) with $p$  either $[0,\infty]$ or $[0,-\infty]$ ensures the orthogonality of time, 
$\bk{t}{t'}=\delta(t-t')$:
\begin{equation}\label{ligai2e}
\bk{p}{t}=\int\bk{p}{E}\bk{E}{t}dE=\frac{1}{\sqrt{2\pi}}\int\sqrt{\frac{p}{E_p}}\delta(\pm E_p-E) e^{iEt} dE
=\frac{1}{\sqrt{2\pi}}\sqrt{\frac{p}{E_p}}e^{\pm i E_p t}
\end{equation}
\begin{align}
\bk{t}{t'} &=\int^\infty_0\bk{t}{p}\bk{p}{t'}dp=\frac{1}{2\pi}\int^\infty_0\frac{p}{E_p}e^{\pm iE_p(t'-t)}dp\\
\comm{&=\frac{1}{2\pi}\int^\infty_0\frac{p}{E_p}e^{iE_p(t'-t)}\frac{E_p}{p}dE_p
+\frac{1}{2\pi}\int^\infty_0\frac{p}{E_p}e^{-ip_E(t'-t)}\frac{E_p}{p}dE_p\\
&=\frac{1}{2\pi}\int^\infty_0 e^{iE_p(t'-t)}dE_p+\frac{1}{2\pi}\int^{-\infty}_0 e^{iu(t'-t)}d(-u)\\}
&=\frac{1}{2\pi}\int^\infty_{-\infty} e^{iE_p(t'-t)}dE_p=\delta(t-t')
\end{align}
If we did not restrict $p$ to either positive or negative values, we couldn't have $\bk{t}{t'}=\delta(t-t')$.
This is expected because the negative momentum particle comes backward in space to be detected at another time 
at the same position it has already been detected.

\section{Application} 
We have seen how to treat the position and time equally in quantum mechanics especially in evolution process. Let's consider some application of our prescrition. 
%The question is in what situation we can apply the quantum evolution by $\underline{x}$. Of course the answer would be the situation where the role of space and time flips. 
\subsection{Time-of-arrival} 
We may apply new prescription to the time-of-arrival introduced earlier. By putting (\ref{ligai2e}) into (\ref{ligah4b1}), we can derive the expression of time-of-arrival for a free particle.
Then (\ref{ligah4b1}) turns out
\comm{
\[
\bk{p}{t}=\frac{1}{\sqrt{2\pi}}\sqrt{\frac{p}{E_p}}e^{\pm i E_p t} \tag{\ref{ligai2e}}
\]
\[
\varphi_2(t_2,\underline{x_2}) = \int \langle t_2\mid p\rangle \varphi_1(p)
e^{ip(x_2-x_1)}dp \tag{\ref{ligah4b1}}
\]
}
\begin{equation}\label{ligah4b5}
\varphi_2(t)=\frac{1}{\sqrt{2\pi}}\int\sqrt{\frac{p}{E_p}}
e^{\mp iE_p t+ip(x_2-x_1)}\varphi_1(p)dp
\end{equation}
where the subscript 1 and 2 in $\varphi$ stands for the position $x_1$ and $x_2$.
The range of $p$ goes either $[0,\infty]$ or $[0,-\infty]$; 
$\mp E_p$ correspond positive and negative energy particle respectively. The negative energy
particle evolves in opposite direction to the positive energy particle in time.  
(\ref{ligah4b5}) is well consistent with the final result Grot \emph{et al}\cite{grot} derived for a free particle.
\\

We have drived (\ref{ligah4b5}) from $\bk{t_2,\underline{x_2}}{p_n}\bk{p_n}{\varphi_1,\underline{x_1}}$. We could also derive (\ref{ligah4b5}) from $\bk{t_2,\underline{x_2}}{t_i,\underline{x_1}}\bk{t_i,\underline{x_1}}{\varphi_1,\underline{x_1}}$.
\begin{align}
\bk{t_2,\underline{x_2}}{\varphi_1,\underline{x_1}} &=\int dt_1\bk{t_2,\underline{x_2}}{t_1,\underline{x_1}}\bk{t_1,\underline{x_1}}{\varphi_1,\underline{x_1}}\label{ligah4b5a}\\
\bk{t_2}{\varphi_2,\underline{x_2}}&= \int dt_1\bk{t_2,\underline{x_2}}{p_m}\bk{p_m}{t_1,\underline{x_1}}\bk{t_1,\underline{x_1}}{p_n}\bk{p_n}{\varphi_1,\underline{x_1}}\\
\varphi_2(t_2)&= \int dt_1\bk{t_2}{p_m}\bk{p_m}{t_1}\bk{\underline{x_2}}{p_m}\bk{p_m}{\underline{x_1}}\bk{t_1}{p_n}\bk{p_n}{\varphi_1}\label{ligah4b5a1}
\end{align}
\begin{align}
\varphi_2(t_2) &=\iiint dt_1\Big(\frac{p_m}{2\pi E_{p_m}}\Big)\sqrt{\frac{p_n}{2\pi E_{p_n}}}
e^{\pm i(E_{p_m}-E_{p_n})t_1}e^{\mp iE_{p_m}t_2} e^{ip_m(x_2-x_1)}\varphi_1(p_n)dp_m dp_n\label{ligah4b5b}\\
&= \iint \frac{p_m}{E_{p_m}}\sqrt{\frac{p_n}{2\pi E_{p_n}}}\delta(E_{p_m}-E_{p_n})e^{\mp E_{p_m}t_2+ip_m(x_2-x_1)}\varphi_1(p_n)\frac{E_{p_m}dE_{p_m}}{p_m}dp_n\label{ligah4b5c}\\
&= \int^\infty_0 \sqrt{\frac{p_n}{2\pi E_{p_n}}} e^{\mp E_{p_n}t_2+ip_n(x_2-x_1)}\varphi_1(p_n)dp_n\label{ligah4b5d}
\end{align}
which reduce to (\ref{ligah4b5}).
\comm{We just check that $\bk{t_2,\underline{x_2}}{\varphi_1,\underline{x_1}}=\bk{t_2,\underline{x_2}}{p_n}\bk{p_n}{\varphi_1,\underline{x_1}}=\bk{t_2,\underline{x_2}}{t_i,\underline{x_1}}\bk{t_i,\underline{x_1}}{\varphi_1,\underline{x_1}}$.}
Note that if we did not restrict the momentum to either $[0,\infty]$ or $[0,-\infty]$, we could not have $e^{ip_n(x_2-x_1)}$ in (\ref{ligah4b5d}) from (\ref{ligah4b5c}). 

\comm{
In classical limit, (\ref{ligai2e}) reduces to
\begin{equation}\label{ligah4b5e}
\bk{p}{t}=\frac{1}{\sqrt{2\pi}}\sqrt{\frac{p}{m}}e^{\pm i\frac{p^2}{2m}t}
\end{equation}
and (\ref{ligah4b5}) reduces to
\begin{equation}\label{ligah4b5f}
\varphi_2(t)=\frac{1}{\sqrt{2\pi}}\int\sqrt{\frac{p}{m}}
e^{\mp i\frac{p^2}{2m} t+ip(x_2-x_1)}\varphi_1(p)dp
\end{equation}
We could also derive (\ref{ligah4b5f}) from $\bk{t_2,\underline{x_2}}{t_i,\underline{x_1}}\bk{t_i,\underline{x_1}}{\varphi_1,\underline{x_1}}$.
By putting (\ref{ligah4b5e}) into (\ref{ligah4b5a1}),
\begin{equation}\label{ligah4b5g}
\varphi_2(t_2)=\iiint dt_1\Big(\frac{1}{2\pi}\frac{p_m}{m}\Big)^{3/2} 
e^{\pm \frac{i}{2m}(p^2_m-p^2_n)t_1}e^{\mp\frac{i}{2m}p^2_m t_2+ip_m(x_2-x_1)}\varphi_1(p_n)dp_n dp_m
\end{equation}
\begin{equation}\label{ligah4b5h}
\int e^{\pm i(p^2_m-p^2_n)\frac{t_1}{2m}}(2m)d\Big(\frac{t_1}{2m}\Big)
=2\pi\cdot 2m\delta(p^2_m-p^2_n)=\frac{4\pi m}{2|p_m|}[\delta(p_m-p_n)+\delta(p_m+p_n)]
\end{equation}
\begin{equation}\label{ligah4b5i}
\varphi_2(t_2)=\iint \Big(\frac{1}{2\pi}\frac{p_m}{m}\Big)^{3/2}
\frac{2\pi m}{|p_m|}[\delta(p_m-p_n)+\delta(p_m+p_n)]
e^{\mp\frac{i}{2m}p^2_m t_2+ip_m(x_2-x_1)}\varphi_1(p_n)dp_n dp_m
\end{equation}
(\ref{ligah4b5i}) reduces to (\ref{ligah4b5f}) only for the same sign of $p_n$ and $p_m$.  In other words, in order to make the prescription self consistent for $\underline{x_2}\neq \underline{x_1}$ in PEP, we need to constraint the range of momentum to either $[0,\infty]$ or $[0,-\infty]$.

We can also check that (\ref{ligaj1a}) cannot make (\ref{ligah4b5i}) and (\ref{ligah4b5f}) consistent each other.
}

\subsection{Quantum tunneling} 
Another application  is the region of quantum tunneling. (Or inside event horizon.) Thus let's apply the prescription to derive quantum tunneling formula.
For $E_1=E_2$ (\ref{ligah8h}) becomes
\begin{align*}
\langle E_1, \underline{x_2}\mid E_1,\underline{x_1}\rangle &= \iint \langle t_2,\underline{x_2}\mid t_1,\underline{x_1}\rangle \frac{1}{2\pi}e^{iE_1(t_2-t_1)}dt_1 dt_2\\
&= \frac{1}{2\pi}\int^\infty_{-\infty}\int^\infty_{-\infty}\sum_{[x(t)]}\exp\Big(i\int^{t_2}_{t_1}(L+E_1)dt\Big) dt_1 dt_2
\end{align*}
where we have used Feynman kernal.
\comm{
\begin{equation}\label{ligah9f} 
\langle t_2,\underline{x_2}\mid t_1,\underline{x_1}\rangle = \sum_{[x(t)]}e^{i\int^{t_2}_{t_1}Ldt}
\end{equation}
}
And we can make it simpler by
\begin{equation}\label{ligah9g} 
\int^{t_2}_{t_1}(L+E_1)dt=\int^{t_2}_{t_1} p\dot{x}dt=\int^{x_2}_{x_1}pdx\equiv W(x)\Big|^{x_2}_{x_1}
\end{equation}
where $p\equiv \frac{\partial L}{\partial \dot{x}}$ and $W$ stand for the generalized momentum and the Jacobi action respectively. Then finally we have
\begin{equation}\label{ligah9h} 
\langle E_1, \underline{x_2}\mid E_1,\underline{x_1}\rangle =\frac{1}{2\pi}\int^\infty_{-\infty}\int^\infty_{-\infty}\sum_{[x(t)]} \exp \frac{iW(x)\Big|^{x_2}_{x_1}}{\hbar} dt_1dt_2
\end{equation}
For a classical object ($W\gg\hbar$) or for WKB approximation \cite{hibbs},
\begin{equation}\label{ligah9i} 
\sum_{[x(t)]} \exp\Big(\frac{iW(x)\Big|^{x_2}_{x_1}}{\hbar}\Big)\sim \exp\Big(\frac{iW(x_\ell)\Big|^{x_2}_{x_1}}{\hbar}\Big)F(t_2,t_1)
\end{equation}
where $F(t_2,t_1)$ is some function of only $t_2$ and $t_1$.  $x_\ell$ stands for the least action path satisfying Euler-Lagrange equation
\begin{equation}\label{ligah9j} 
\frac{\partial}{\partial t}\Big(\frac{\partial L}{\partial \dot{x}}\Big)-\frac{\partial L}{\partial x}=0
\end{equation}
Thus
\begin{equation}\label{ligah9k} 
\langle E_1, \underline{x_2}\mid E_1,\underline{x_1}\rangle  \simeq \exp\Big(\frac{i}{\hbar}W(x_\ell)\Big|^{x_2}_{x_1}\Big)
\end{equation}
For tunneling particle, $p^2<0$, $W=i(\textrm{Im}W)$,
\begin{equation}\label{ligah9l} 
\langle E_1, \underline{x_2}\mid E_1,\underline{x_1}\rangle \sim e^{-\frac{1}{\hbar}\textrm{Im}W(x_\ell)\Big|^{x_2}_{x_1}}
\end{equation}
The tunneling probability is
\begin{equation}\label{ligah9m} 
P(\underline{x_1}\to \underline{x_2},E)\sim e^{-\frac{2}{\hbar}\textrm{Im}W(x_\ell)\Big|^{x_2}_{x_1}}
\end{equation}

In (\ref{ligah6c}), we have seen that a particle in energy eigenstate departs the initial position $\underline{x_1}$ at all different times equally. This applies
also to the final position $\underline{x_2}$ in (\ref{ligah8h}). Thus it is meaningless to talk about tunneling time of a particle in energy eigenstate;
(\ref{ligah4b5}) reveals this property clearly. We can consider (\ref{ligah4b5}) as tunneling time for a zero potential, $V=0$. For an energy eigenstate 
$\varphi_1=p_1$, $|\varphi_2(t)|^2$ of (\ref{ligah4b5}) has no time dependence. It makes sense, because the particle in energy eigenstate departs $x_1$ and
arrives $x_2$ at all different times equally. We may discuss time-of-arrival or tunneling time only for particles which are not in energy eigenstate.

\comm{
\section{MN}

We can pass the evolution parameter to the basis by
\begin{equation}\label{ligah3c} 
\mid \psi_1,\underline{t_1}\rangle = \sum_n \mid\phi_n,\underline{t_1}\rangle\langle \phi_n,\underline{t_1}\mid \psi_1, t_1\rangle =\sum_n \mid \phi_n,\underline{t_1}\rangle\langle \phi_n\mid \psi_1\rangle
\end{equation}

We can pass the evolution parameter to the basis by
\begin{equation}\label{ligah4c} 
\mid \varphi_1,\underline{x_1}\rangle = \sum_n \mid\phi_n,\underline{x_1}\rangle\langle \phi_n,\underline{x_1}\mid \varphi_1, x_1\rangle =\sum_n \mid \phi_n,\underline{x_1}\rangle\langle \phi_n\mid \varphi_1\rangle
\end{equation} 
\\

From (\ref{ligah8h}), 
\begin{eqnarray*}
\langle E_2,\underline{x_2}&\mid& E_1,\underline{x_1}\rangle \\ 
&=&\int^\infty_{-\infty}\int^\infty_{-\infty}\langle t_2,\underline{x_2}\mid t_1,\underline{x_1}\rangle \frac{1}{2\pi}e^{iE_2t_2-iE_1t_1}dt_2dt_1 \\
&=& \int^\infty_{-\infty}\int^\infty_{-\infty} \sum_n\langle t_2,\underline{x_2}\mid E_n\rangle\langle E_n\mid t_1,\underline{x_1}\rangle \frac{1}{2\pi}dt_2dt_1\\
&=&\int^\infty_{-\infty}\int^\infty_{-\infty}\sum_n\frac{1}{4\pi^2}e^{i(E_2-E_n)t_2}e^{i(E_n-E_1)t_1}dt_1dt_2
\end{eqnarray*}
You may think this is zero unless $E_1=E_2$ and $\underline{x_2}$ is implied in $t_2$. But this is wrong because you sandwiched $\mid E_n\rangle\langle E_n\mid$ between $\underline{x_2}$ and $\underline{x_1}$.
\\

Could any operator  be evolution parameter?
\[
\mid \psi_1,\underline{\phi_1}\rangle=
\]
We used Feynman kernel $\langle x_2,\underline{t_2}\mid x_1,\underline{t_1}\rangle$ for $\langle t_2,\underline{x_2}\mid t_1,\underline{x_1}\rangle$. How could it be possible? If it is not possible how can we calculate $\langle t_2,\underline{x_2}\mid t_1,\underline{x_1}\rangle$ in space-time symmetric way by
\[
\langle t_f,q_f\mid t_i,q_i+q\rangle=\langle t_f,q_f\mid e^{-i\hat{p}q}\mid t_i, q_i\rangle
\] 

It seems the eigenstate of $\hat{T}$ do not orthogonal each other.

The quantum tunneling formula (\ref{ligah9m}) derived from (\ref{ligah8h}) illustrated similar to figure \ref{tepeptun_fig2}(b). Some arrows move between $x_1$ and $x_2$ faster than the speed of light and some arrows move even backward in time. However this does not mean the particle actually move faster than $c$ or move backward in time: The stationary particle of mass $m$  at $x_0<x_1$ described by the Gaussian wave packet with uncertainty $\Delta x_0$ evolves to the wave packet centered at $x_0$ with uncertainty $\Delta x_1$ after $\Delta t$
\[
\Delta x_1=\Delta x_0\sqrt{1+\Big(\frac{\hbar \Delta t}{m\Delta x^2_0}\Big)^2}
\]
If the particle has smaller mass, there would be more probability to be found at $x_1$. This probability provides the probability function $x_1$ in figure \ref{tepeptun_fig2}(b)....
}

\section{Conclusion}
We have seen how to put an equal footing on position and time in quantum mechanics. Unlike other approaches, I proposed that we cannot take both position and time as evolution parameters or both as observables. We have to take one as an observable and the other as an evolution parameter; With set of simple prescriptions, we could formulate quantum mechanics in space time symmetric manner. Combining with Feynman path integral, we could understand the fundamental meaning of time-energy uncertainty principle. 
We could derive the time-of-arrival for a free particle.
We could also develop quantum tunneling formula expressed in Jacobi action for classical or WKB limit.
This approach may contribute to the development of quantum gravity.

One drawback of this prescription is that, for example figure \ref{tepeptun_fig2}(b) suggest that the particle can travel faster than the speed of light or even backward in time. We can fix this problem by just assuming  it cannot do it and modifying the integral range in formula for final position. But this is not an elegant way to bypass the problem and it ruins the spirit of space time symmetry we are trying to achieve. Does that mean the prescription for the position as an evolution parameter applies only to stationary case where there is no measurable distinction between past and future? Further study is needed to answer it.

\section*{Acknowledgments}
I would like to thank Werner Israel for his support and useful comments on this work.

\appendix
\section{Inconsistency of making $\bk{E}{p}$ even function}\label{epeven}
We saw that (\ref{ligai2c}) is an odd function of $p$. In order to achieve
$\bk{E'}{E}=\delta(E'-E)$ we had to restrict $p$ to either $[0,\infty]$ or $[0,-\infty]$. However, we may achieve it by choosing
\begin{equation}\label{ligaj1a}
\bk{E}{p}=\sqrt{\frac{|p|}{2E_p}}\delta(\pm E_p-E) 
\end{equation}
\begin{equation}\label{ligaj1b}
\bk{p}{t}=\int\bk{p}{E}\bk{E}{t}dE=\int^\infty_{-\infty}\sqrt{\frac{|p|}{2E_p}}\delta(\pm E_p-E)\frac{e^{iEt}}{\sqrt{2\pi}}dE=\frac{1}{\sqrt{2\pi}}\sqrt{\frac{|p|}{2E_p}}e^{\pm iE_pt}
\end{equation}
Then, we can show that (\ref{ligaj1a}) satisfies $\bk{E'}{E}=\delta(E'-E)$ and
$\bk{t'}{t}=\delta(t'-t)$ without constraining allowed range of momentum for the same evolution parameter $\underline{x_1}$.
\comm{
\begin{align*}
\bk{E'}{E} &= \int^\infty_{-\infty}\bk{E'}{p}\bk{p}{E}dp
=\int^\infty_{-\infty}\frac{|p|}{2E_p}\delta(\pm E_p-E')\delta(\pm E_p-E)dp\\
&=\int^\infty_0\frac{|p|}{2E_p}\delta(\pm E_p-E')\delta(\pm E_p-E)dp
+\int^0_{-\infty}\frac{|p|}{2E_p}\delta(\pm E_p-E')\delta(\pm E_p-E)\underbrace{dp}_{\frac{E_p dE_p}{p}}\\
&=\int^\infty_0\frac{1}{2}\delta(\pm E_p-E')\delta(\pm E_p-E)dE_p
+\int^0_\infty \frac{1}{2}(-1)\delta(\pm E_p-E')\delta(\pm E_p-E)dE_p\\
&=\int^\infty_0\delta(\pm E_p-E')\delta(\pm E_p-E)dE_p
=\int^\infty_{-\infty}\delta(u-E')\delta(u-E)du\\
&=\delta(E'-E)
\end{align*}
\begin{align*}
\bk{t'}{t} &=\int^\infty_{-\infty}\bk{t'}{p}\bk{p}{t}dp
=\int^\infty_{-\infty}\frac{1}{2\pi}\frac{|p|}{2E_p}e^{\pm iE_p(t-t')}dp\\
&=\int^\infty_0 \frac{1}{2\pi}\frac{|p|}{2E_p}e^{\pm iE_p(t-t')}dp
+\int^0_{-\infty} \frac{1}{2\pi}\frac{|p|}{2E_p}e^{\pm iE_p(t-t')}\underbrace{dp}_{\frac{E_p dE_p}{p}}\\
&=\int^\infty_0\frac{1}{4\pi}e^{\pm iE_p(t-t')}dE_p
+\int^0_\infty\frac{1}{4\pi}(-1)e^{\pm iE_p(t-t')}dE_p\\
&=\int^\infty_0\frac{1}{2\pi}e^{\pm iE_p(t-t')}dE_p
=\int^\infty_{-\infty}\frac{1}{2\pi}e^{iu(t-t')}du\\
&=\delta(t'-t)
\end{align*}
}
However, we can also show that (\ref{ligaj1a}) does not work consistently for
$\bk{t_2,\underline{x_2}}{\varphi_1,\underline{x_1}}$.
Using (\ref{ligaj1b}) to (\ref{ligah4b1}), we have
\begin{equation}\label{ligaj1g}
\varphi_2(t)=\frac{1}{\sqrt{2\pi}}\int^\infty_{-\infty}\sqrt{\frac{|p|}{2E_p}}
e^{\mp iE_p t+ip(x_2-x_1)}\varphi_1(p)dp
\end{equation}
and (\ref{ligah4b5c}) changes to
\begin{align}
\varphi_2(t_2) &= \int^\infty_{-\infty}\int^\infty_0 \frac{1}{2}\sqrt{\frac{|p_n|}{2\pi 2E_{p_n}}}\delta(E_{p_m}-E_{p_n})e^{\mp iE_{p_m}t_2+ip_m(x_2-x_1)}\varphi_1(p_n)dE_{p_m} dp_n\\
&+\int^\infty_{-\infty}\int^0_\infty -\frac{1}{2}\sqrt{\frac{|p_n|}{2\pi 2E_{p_n}}}\delta(E_{p_m}-E_{p_n})e^{\mp iE_{p_m}t_2+ip_m(x_2-x_1)}\varphi_1(p_n)dE_{p_m} dp_n\\
&= \int^\infty_{-\infty}\int^\infty_0 \sqrt{\frac{|p_n|}{2\pi 2E_{p_n}}}\delta(E_{p_m}-E_{p_n})e^{\mp iE_{p_m}t_2+ip_m(x_2-x_1)}\varphi_1(p_n)dE_{p_m} dp_n\label{ligaj1h}
\end{align}
(\ref{ligaj1h}) is not consistent with (\ref{ligaj1g}) unless $\underline{x_2}=\underline{x_1}$ or $p_m$ and $p_n$ have the same sign. 
We can also check this inconsistency  in classical limit.

\end{document}